# Avoiding the Great Filter: Predicting the Timeline for Humanity to Reach Kardashev Type I Civilization

Jonathan H. Jiang[1], Fuyang Feng[2], Philip E. Rosen[3], Kristen A. Fahy[1],
Antong Zhang[4], Piotr Obacz[5], Prithwis Das[6], Zong-Hong Zhu[2]

[1]Jet Propulsion Laboratory, California Institute of Technology, Pasadena, California, USA

[2]Department of Astronomy, Beijing Normal University, Beijing, China

[3]Retired, Chevron Energy Technology Company, Houston, Texas, USA

[4]Damien High School, La Verne, California, USA

[5]Faculty of International and Political Studies, Jagiellonian University, Kraków, Poland

[6]Vivekananda Mission High School, India

Correspondence: Jonathan.H.Jiang@jpl.nasa.gov

## Abstract

The level of technological development of any civilization can be gaged in large part by the amount of energy they produce for their use, but also encompasses that civilization's stewardship of their home world. Following the Kardashev definition, a Type I civilization is able to store and use all the energy available on its planet. In this study, we develop a model based on Carl Sagan's K formula and use this model to analyze the consumption and energy supply of the three most important energy sources: fossil fuels (e.g., coal, oil, natural gas, crude, NGL and feedstocks), nuclear energy and renewable energy. We also consider environmental limitations suggested by United Nations Framework Convention on Climate Change, the International Energy Agency, and those specific to our calculations to predict when humanity will reach the level of a Kardashev scale Type I civilization. Our findings suggest that the best estimate for this day will not come until year 2371.

## 1. Introduction
### 1.1 The Definition and Classification of Kardashev Scale

As a method to measure and classify the technological advancement of a civilization, Soviet astronomer Nikola Kardashev conceived of the Kardashev Scale in 1964 [1]. In its early formulation, the Kardashev Scale was based on the overall consumption of energy (assumed equal to the total energy supply) of a given civilization. Kardashev defined 3 distinct types of civilizations.

A Type I civilization is referred to as a "planetary civilization". Characterized as having a power consumption of $10^{16}$ W (Watts) [1,2], a civilization of Type I has harnessed for its use all major forms of energy available from its home planet (for example human civilization and the Earth) and also includes the energy received by the home world from its parent star. Planetary energy sources humanity can tap include fossil and bio-derived fuels, nuclear energy, wind, solar, geothermal, and tidal, among others.

A Type II civilization, also referred to as a "stellar civilization", consumes $10^{26}$W [2]. To achieve ten orders of magnitude increase in power consumption over Type I civilizations, a Type II civilization must be able to obtain and store all the energy its parent star releases. Of course, a Type II civilization can also take energy from its planetary system as well. To achieve such a colossal degree of energy capture and consumption, immense constructions would all but certainly need to be engineered and built. For example, a Dyson sphere [3,4], Matryoshka Brain [5,6,7] or other forms of planetary system networks or star lift (also called stellar mining, stellar engineering, or astro-engineering [8,9,10].) The sun's total energy output, its luminosity, is approximately $4 \times 10^{26}$W [2]. If human civilization can capture more than one fourth of the sun's overall energy output and use it efficiently, ours will qualify as a Type II civilization. Other, more exotic energy sources of Type II civilizations may include extraction from black hole accretion disks and/or jets [11] or matter-antimatter annihilation. As an approximation to the more familiar, it can be estimated that to obtain $1 \times 10^{26}$W, the water mass contained in approximately 14,700 typical backyard swimming pools, typically 20,000 gallons each, would need to be annihilated every second, assuming 100% efficiency in converting mass to captured energy.

A Type III civilization is also called a "galactic civilization". With an energy consumption of $10^{36}$W [2], a civilization of Type III can access and control much of the energy the entire galaxy that civilization lives in generates. Considering the luminosity of the entire Milky Way galaxy is about $4 \times 10^{37}$W [2], a Type III civilization residing in humanity's home galaxy would command minimally 2.5% of the total possible energy sources, a notion which stretches to the breaking point even the most highly theoretical means to do so – white holes, supermassive black holes or some sources human civilization is not able to even conceive yet. Indeed, such a civilization may well be able to manipulate space-time itself, creating wormholes to enable travel to any point in the Universe.

**1.2 Carl Sagan's Formulation of the Continuous Kardashev Scale**

To address the quantitative limitations inherent in the discretized original definition of the Kardashev Scale, American astronomer Carl Sagan formulated a definition using the whole energy power P, in watts, in a logarithm calculation which yielded a continuous function expressing the Kardashev Scale, K [12]:

$$K = \frac{lg(P) - 6}{10} \qquad (1.1)$$

By using the Equation 1.1, we can calculate the power of Type I, II and III civilizations easily, as well as extrapolate below Type I to quantify civilizations which command only a very small fraction of their world's available energy. According to data from the IEA (International Energy Agency), in 2018 the total energy supply of the entire world was 1,4281,889 kTOE (an energy unit called kilo ton oil equivalent), or approximately $1.90 \times 10^{13}$W. From this actual human civilization power production estimate and Eq. 1.1, we can calculate the present value of K for human civilization to be 0.728.

Of more compelling interest than where precisely humanity presently resides on the Kardashev

Scale is when our species will reach Type I civilization status. Some researchers have attempted to calculate such a timeline. In his article [13], Michio Kaku concluded if human growth averages at a rate of about 3% per year, we will reach Type I in 100~200 years, Type II possibly in a few thousand years, and Type III in, perhaps, 0.1~1 million years. A recent study [14] has calculated the timeline for humanity attaining Type I status to be approximately the year 2347. However, this estimate is the result of a simple exponential growth model for calculating total energy production and consumption as a function of time, relying on a continuous feedback loop and absent detailed consideration of practical limitations. With this reservation in mind, its prediction for when humanity will reach Type I civilization status must be regarded as both overly simplified and somewhat optimistic. Accordingly, our modeling will consider the following three main sources of the energy and their associated limitations. These sources include fossil fuels (conventional reservoir oil, coal, peat, oil shale, natural gas, crude oil, NGL and feedstocks), nuclear and renewable energy. As well, we will consider the policies and suggestions from the United Nations Framework Convention on Climate Change (UNFCCC) and the IEA's forecasts for energy consumption in the several decades to come to aid in developing our model for estimating the timeframe for humanity to attain Kardashev Type I status.

Technological development over the past 5,000 years of human civilization has led our species to dominance of life on Earth and placed us on a pathway to achieving a Type I civilization – and perhaps beyond. However, coupled with continuing and profound sociological challenges our progress also threatens to end or severely set back our civilization. As we face what has been described as the "Great Filter" [15], technology can also hold the key to our long-term survival. Some have proposed leveraging rapidly advancing technical capability to establish robust off-world colonies, carrying forward humanity's legacy into the indefinite future should global disaster befall our home world [16]. Together with such lofty plans, attaining a Type I civilization would logically find humanity having solved many of the most vexing problems which have beset us since invention of the first primitive written languages set us upon our current path. Apart from the uncounted millions of Earth's species past and present, technological progress has placed in humanity's hands the future of our world and every living creature upon it. Clearly, there can be no turning away from further advancement. How we choose to proceed along that arc is of the upmost importance and urgency.

## 2.0 Methodology and Calculations
### 2.1 Fossil Fuels

Fossil fuels, as an essential energy source of our modern society, play an important role in our technological development such as generating electricity, cooking, manufacturing petrochemical products and as the main fuel for nearly all modes of transportation. However useful, we must accept that the production and burning of hydrocarbons carries with it highly impactful environmental downsides. Specifically, this takes the form of an increasingly influential greenhouse effect, a result primarily of $CO_2$ emissions to the atmosphere, which is driving up global average temperature. Recognizing this threat, the UNFCCC has recommended that "*Aggregate emission pathways consistent with holding the increase in the global average* *temperature to well below 2°C above preindustrial levels and pursuing efforts to limit the temperature increase to 1.5°C above preindustrial levels.*" Further, the IEA has suggested $CO_2$ emissions worldwide must reach net zero by 2050 to achieve this dramatic energy production and usage transition during these next three decades. To reduce $CO_2$ emissions and slow the increase of global temperatures, a number of countries and organizations have stepped forward with (non-binding) commitments. China, as the second largest economy in the world, has stated their $CO_2$ emissions will reach a peak in 2030 and, declining from then, accomplish carbon neutrality not later than 2060 [25]. Similarly, the European Union (EU) has claimed that "*The binding Union 2030 climate target shall be a domestic reduction of net greenhouse gas emissions (emissions after deduction of removals) by at least 55 % compared to 1990 levels by 2030.*" [17].

### 2.1.1 Uninfluenced Model for Fossil Fuels

In this section we will consider data from the IEA on coal, natural gas, crude, NGL (Natural Gas Liquid) and feedstocks usage between 1971 and 2019, but will initially leave aside any influence originating from policies or related suggestions of the UNFCCC and IEA. However, it is important to point out in IEA's definition "Natural gas includes both 'associated' and 'non-associated' gas as well as colliery gas (excluding natural gas liquids)." Starting from an assumption of energy consumption following an annually compounded curve, a simple growth model emerges to predict consumption values into the future. Coupled with data from the IEA, we can calculate the annual growth rate $t_n$ by Eq 2.1:

$$t_n = \frac{a_{n+1} - a_n}{a_n} \qquad (2.1)$$

where $a_n$ represents energy consumption for the nth year starting from the year of 1971. Using this formulation for $t_n$, we next obtain the average growth rate of consumption $t_{increase}$ by the following Eq 2.2:

$$t_{increase} = \bar{t}_n = \frac{1}{n}\sum_{i=1}^{n} t_i \qquad (2.2)$$

After that, we can calculate the uninfluenced value of the energy consumption by Eq 2.3:

$$y_{un\_inc}(x) = a_{n_0}(1 + t_{increase})^x \qquad (2.3)$$

where x is the number of years after 1971.

### 2.1.2 Influenced Model for Fossil Fuels

By fully factoring in some of the more significant influences from government and organization policies, the consumption of fossil fuels would reach their peak in 2030 and achieve a net zero in 2050. Thus, from now to 2030, the consumption value will obey the classical exponential model discussed in section 2.1.1. Further, we can restrict the range of the years to between 1971 and 2030 with the following Eq 2.4:

$$y_{influence}(x) = a_{1971}(1 + t_{increase})^x, x \in [0,59] \tag{2.4}$$

where the subscript "*influence*" denotes the influenced portion of the overall increase in energy consumption, "$a_{1971}$" is the initial consumption in 1971, and $x$ represents the number of years having elapsed after 1971 ranging to 59 which corresponds to the final year of 2030. By using the exponential growth model, the initial energy consumption value $a_{1971}$ and the averaged growth rate of $t_{increase}$, we can forecast energy consumption through the 2020s.

For the decreasing portion of the curve, which is expected to occur over the years 2030 to 2050, we use a similar form to Eq 2.3 as described by the classical decay Eq 2.5 and a modified exponential decay model to replace the classic exponential decay function yielding Eq 2.6:

$$y_{dec}(x) = a_{2030}(1 - t_{decrease})^x \tag{2.5}$$

$$y_{dec}(x) = \prod_{i=0}^{x} a_{2030}[1 - t_0(1 + mi)]^i, x \in [0, 20], x \in \mathbf{N}^* \tag{2.6}$$

where $a_{2030}$ is the consumption value in 2030, $t_0$ is the initial decay rate and $m$ is the rising proportion of the decay rate per year. The modified exponential decay function $y_{dec}(x)$ means the energy consumption value will decrease with an initial decay rate $t_0$ and that decay rate will accelerate by a proportion of $m$ per year.

Next, we will determine the relationship between $t_0$ and $m$. In Eq 2.5 and Eq 2.6, the $t_{decrease}$ and $t_0(1 + mi)$ will influence the consumption $y(x)$ with the initial value $a_{2030}$ and this influence will be from 2030 to 2050 as the independent variable $x$ ranges from 0 to 20. In terms of the classical exponential decay model, Eq 2.5, the influence of every year is the same value $t_{decrease}$. However, in the modified exponential decay model, Eq 2.6, the influence varies from year to year, becoming successively larger as the factor $m$ rises. To balance the influence of these two fitting functions and determine the constrained relation between $t_0$ and $m$, we then integrate over the 20 years from 2030 to 2050:

$$\int_0^{20} t_{decrease}\, di = \int_0^{20} t_0(1 + mi)\, di \tag{2.7}$$

Simplifying, we get the constrained form between the two parameters:

$$t_{decrease} = t_0(1 + 10m) \tag{2.8}$$

We then assume the energy consumption in 2050 is equal to 1% of the 2030 value, since some aircraft, most oceangoing surface ships and some other means of transportation maybe still be utilizing hydrocarbon-based fuel as their main energy source. So, we make a relatively conservative assumption – 1% (or 2%). And further, we accept as a given that the assumption we have mentioned above will satisfy the carbon dioxide emissions net zero in 2050 requirement. Note that this assumption may be challenged given that some indeterminant hydrocarbon fuel production and usage will likely still be needed after 2050. While application of large-scale atmospheric $CO_2$ removal strategies may help to counterbalance continued (relatively low level) emissions, absent such efforts another way to express this would be to assume $CO_2$ emissions bottom out at some non-zero level starting in 2050 and continue at that level into the indefinite future. This would require a constant be added:

$$a_{2030}(1-t_{decrease})^{20} = 1\% a_{2030} \tag{2.9}$$

The 1% assumption yields a $t_{decrease}$ of 20.57%. Similarly, by redefining to 2%, $t_{decrease}$ declines to 17.77%. By defining the initial decay rate $t_0$, we can get the value of $m$. In our model $t_0$ is

$$t_0 = \frac{1}{10} t_{decrease} \tag{2.10}$$

By Eq 2.8 and applying the constraint defined by Eq 2.10, we determine $m = 0.9$. Hence, an example is generated to describe the annual decay rate from 2030 to 2050 with the Eqs 2.8~2.10.

As can be seen from Table 1, the decay rate in the first decade is relatively low but accelerates in the second decade. Clearly, there is a long way to go in making this dramatic energy transition given that in the 2030s fossil fuels will still play an important part in our total energy supply. As well, within this process of transition some new types of energy may well come to encompass an increasing proportion of our total energy supply.

**Table 1.** Annual decay rate from 2030 to 2050 in our model

| Year | Decay rate | Year | Decay rate |
|---|---|---|---|
| $t_0$ | 2.057% | 2040 | 20.567% |
| 2030 | 2.057% | 2041 | 22.418% |
| 2031 | 3.908% | 2042 | 24.269% |
| 2032 | 5.759% | 2043 | 26.120% |
| 2033 | 7.610% | 2044 | 27.971% |
| 2034 | 9.461% | 2045 | 29.822% |
| 2035 | 11.312% | 2046 | 31.673% |
| 2036 | 13.163% | 2047 | 33.525% |
| 2037 | 15.014% | 2048 | 35.376% |
| 2038 | 16.865% | 2049 | 37.227% |
| 2039 | 18.716% | 2050 | 39.078% |

Ideally, humanity will ultimately free itself from dependence on fossil fuels if governments are sufficiently conscious of the potential environmental problems they pose and alternative energy resources such as nuclear, solar and wind are developed to the extent where they can replace hydrocarbons. As the transition process comes to its assumed end, consumption of fossil fuels will decrease to a relatively low value as our society will by then have only minimal need for it. Given that assumption, it then follows that the decay rate will be relatively large in the 2040s. Here, we have defined just a single example from parameters assumed to have relatively reasonable values. Different assumptions, be them conservative and when used in exponential calculations, will have the potential to yield dramatically different result.

As illustrated in the Figures 1a, b & c, the consumption of the fossil fuels coal, natural gas, crude oil, NGL and feedstocks reach their respective peaks in 2030 before rapidly declining to a relatively low value in 2050. Note that throughout the 20-year period of decrease, the decay rate will increase by a fixed proportion of $m$.

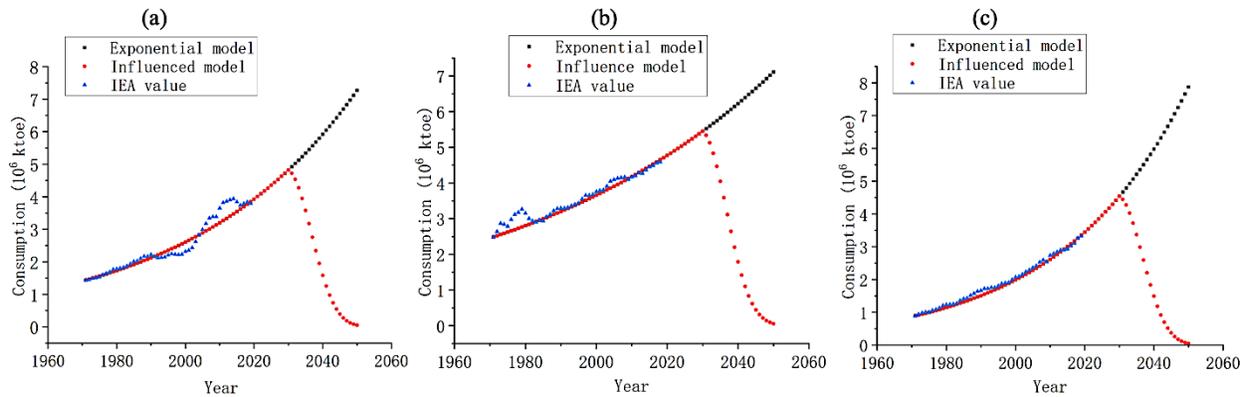

**Figure 1.** The consumptions of (a) coal, (b) natural gas, and (c) crude oil, NGL and feedstocks from 1971 to 2050.

## 2.2 Nuclear and Renewable Energy

"*Nuclear power has historically been one of the largest contributors of carbon-free electricity globally and has significant potential to contribute to power sector decarbonization.*" (IEA). Nuclear power has helped to avoid an estimated 55 Gtons of additional $CO_2$ emissions over the past 50 years, nearly equal to two years of global energy-related $CO_2$ emissions (IEA). Renewables, including solar, wind, hydroelectric, biofuels, and others, are at the center of the transition to a less carbon-intensive and more sustainable energy environment.

"*Biofuels are liquid, solid, or gaseous fuels made from renewable biological materials and are one of the more promising forms of energy for the replacement of non-renewable energy sources (fossil fuels)*" [18]. Comparing with fossil fuel, biofuels are regarded as one important part in reducing $CO_2$ emission in the future. Although biofuels will also increase the content of $CO_2$ in the atmosphere when refined and subsequently oxidized in engines and generators of all types, ultimately, the energy and carbon in biofuel comes from solar energy, soil, and the atmosphere by photosynthesis of plants. Thus, to some extent, burning biofuels will not contribute quite as much $CO_2$ emissions when compared to fossil fuels. As well, biofuels produced by different plants can reduce the $CO_2$ emission in a different proportion [19].

In 2019, renewable energy accounted for 23.2% of global power generation [20]. Clearly, these resources will be an important part of the low to no-carbon society of the future. To achieve full energy transition, the IEA suggests two thirds of all energy production will come from renewable sources including but not limited to solar, wind and geothermal, with the remaining one third covered by nuclear power. According to IEA data, the average growth rate of nuclear is 7.53% from 1971 to 2018. By contrast, growth in renewables is just 2.17%.

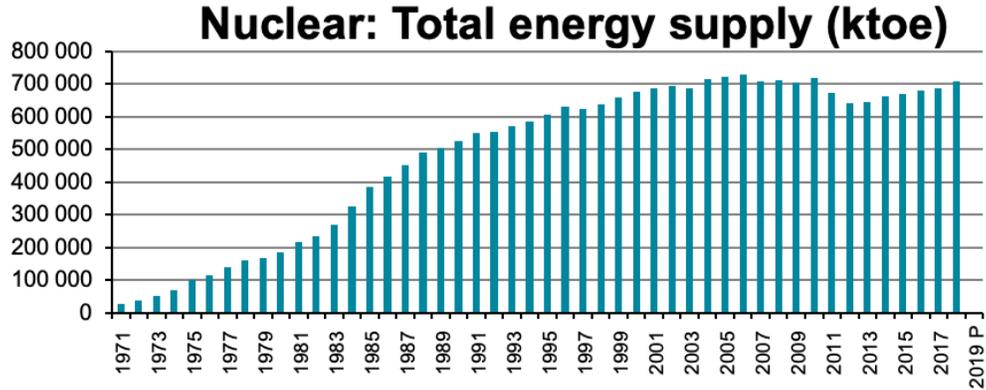

**Figure 2.** The total energy supply of nuclear from 1971 to 2018.

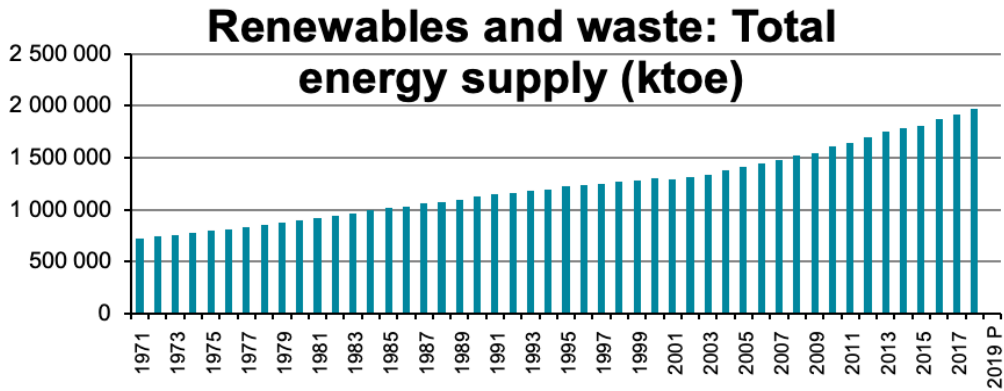

**Figure 3.** The total energy supply of renewables from 1971 to 2018.

Although renewable energy now accounts for a bigger share than nuclear power in the total energy supply picture, to meet the IEA's suggestion we will need to make more of an effort to develop renewables in the coming decades. By considering the IEA's recommendations, we can constrain renewable energy to be two times more than that of nuclear energy:

$$R_{2018}(1+t_{R-increase})^{32} = 2 \cdot N_{2050} \quad (2.11)$$

where, according to the value from 1971 to 2018, we use the classical exponential model and project the values from 2019 to 2050. $R_{2018}$ is the renewable energy supply in 2018, $t_{R-increase}$ is the average growth rate of renewable energy during 2019~2050, and $N_{2050}$ represents the nuclear energy supply in 2050. However, the weaker growth rate for the 20 years between 1998 and 2017 (inclusive) is likely due to slowing of technological advancements relative to prior time periods and/or economic factors. Accordingly, we hold that nuclear energy will not reach that high growth rate of 7.53%, but instead keep to a more moderate rate.

In consideration of the relatively smooth growth data of renewable energy (Figure 3), this as compared with the more fluctuant growth of nuclear (Figure 2), we will regard renewable energy as a standard to estimate the nuclear data accordingly. As in Eq 2.11, $N_{2050}$ is the nuclear energy supply in 2050 and is calculated by the exponential growth model as follows:

$$N_{2050} = N_{2018}(1 + t_{N-increase})^{32} \qquad (2.12)$$

By using Eqs 2.11 and 2.12, $t_{R-increase} = 2.17\%$ and the values of $N_{2018}$ and $R_{2018}$, we can determine $t_{N-increase} = 3.24\%$.

However, according to the estimation from the IEA (Figure 4), under the most ideal condition for the future development of nuclear, this per the Safety Data Sheet (SDS) condition in Figure 4, the average growth rate is about 2.47%. Thus, 3.24% remains too high even when compared with the IEA's estimation of the *idealized* growth rate. Moreover, nuclear energy's average growth rate from 1998 to 2017 was only 0.550%. Furthering the point, from Figure 4 we can make a reasonable prediction that nuclear energy will not grow nearly so fast as 3.24%. In view of this uncertainty, we prudently choose the IEA's estimation of 2.47% as our ideal growth model.

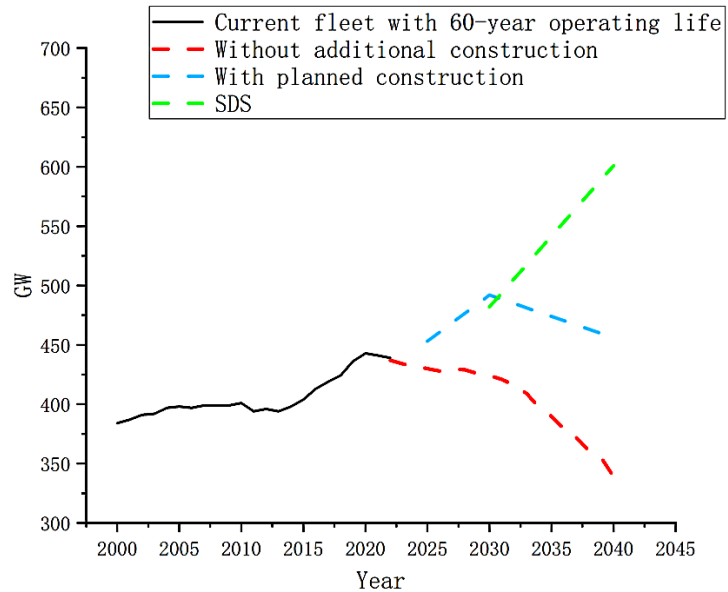

**Figure 4.** The development of nuclear energy under differing conditions.

In this ideal condition, the total energy supply of nuclear in 2050 is calculated by the following equation:

$$N_{2050} = N_{2018}(1 + 2.47\%)^{32} \qquad (2.13)$$

According to the above calculations and considering the IEA's recommendations, we describe in Figure 5 the renewables and nuclear energy outlook within the influenced model. It is evident that to meet the IEA's suggestion we must develop nuclear energy further with more construction while simultaneously maintaining the pace of renewable energy growth.

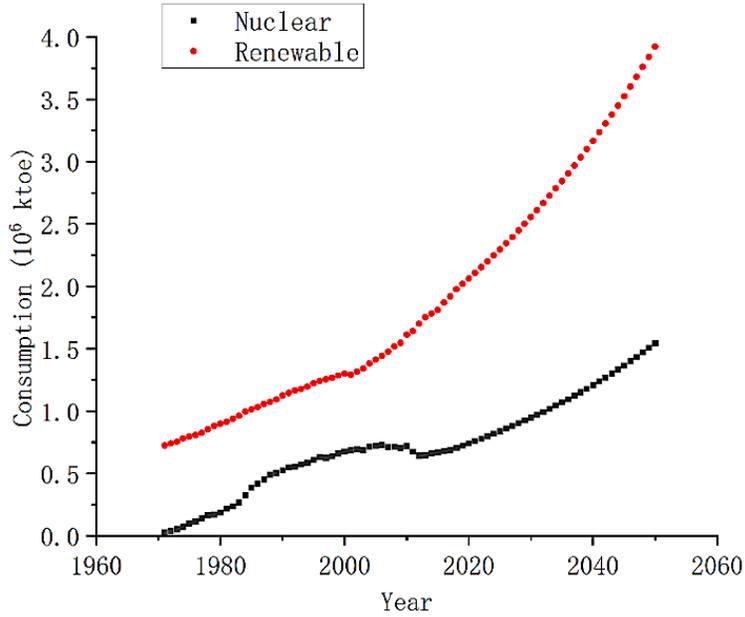

**Figure 5.** The supply of nuclear and renewable energy in the influenced model.

## 3.0 Conclusions

So far, we have discussed the five major types of energy sources (Coal, Natural gas, Crude, Nuclear and Renewable) and analyzed how their usage may change in the near future under the influences and policies raised by the UNFCCC and IEA. Returning to Eq 1.1, Carl Sagan's formula for calculating the Kardashev scale, we project that human civilization can indeed attain a K value of 1.00 with these five energy sources.

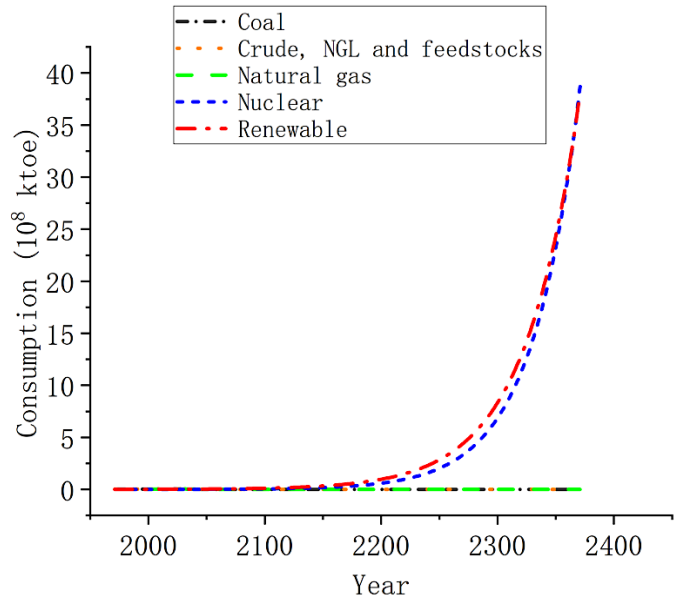

**Figure 6.** The energy supply in the influenced model. Note: Coal is minimal for 1971-2050 and largely coincides with the Natural gas line.

In Figure 6 it is seen that by considering the aforementioned international organizations' suggestions, renewable and nuclear energy will come to take up the largest proportion of total energy supply while fossil fuels will only play a minimal role. If this outcome holds, humanity may well avoid a future beset with the more severe environmental problems posed by excessive $CO_2$ emissions.

A final revisit of Eq 1.1, which is informed by the IEA and UNFCCC's suggestions, finds an imperative for a major transition in energy sourcing worldwide, especially during the 2030s. Although the resultant pace up the Kardashev scale is very low and can even be halted or reversed in the short term, achieving this energy transformation is the optimal path to assuring we will avoid the environmental pitfalls caused by fossil fuels. In short, we will have met the requirements for planetary stewardship while continuing the overall advancement of our technological civilization. Keeping to this developing philosophy, we estimate attainment of Type I civilization status in the year 2371 (Figure 7).

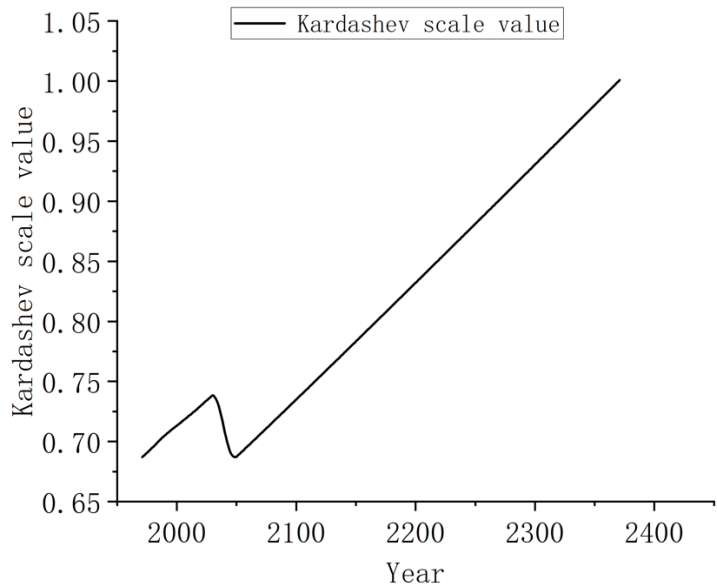

**Figure 7.** The K value in the constrained model.

## 4.0 Discussion and Implications

Our model, which uses the classical exponential growth standard to simulate future energy supplies of nuclear and renewable energy, finds meeting the relatively ambitious average annual growth rate of 2.47% may present a major international challenge. A more pragmatic projection of average annual growth rate may well place it substantially lower than 2.47%. More specifically, this method employs an idealized model, especially for nuclear and renewable energy usage into the future, relying on assumedly benevolent influences from the political, social, and financial realms, the uncertainty of which increases sharply with time. Accordingly, our prediction for humanity's arrival date at a Type I civilization is possibly biased towards the earlier end of the rational range.

There are several difficult to predict reasons affecting the total energy supply as our civilization advances. For example, technological and/or social barriers may arise which inhibit acceptance and utilization of new energy technology. In such circumstances it would be advisable to find new methods of calculation to re-balance the proportion between nuclear energy and renewable energy as we progress.

Despite the contribution from nuclear and the rapid growth in renewables, energy-related $CO_2$ emissions hit a record high in 2018 as electricity demand growth outpaced increases in low-carbon derived power. Humanity should, however, strive to achieve a dramatic energy transition during 2020 to 2050, indeed as soon as possible, to solve at least the most pressing of environmental problems which stand in the path of development to a higher order civilization. Extending this notion further, the stability of a civilization at any Kardashev/Sagan level must necessarily be considered in order to complete a holistic analysis. As an example, if a given civilization produces and consumes a sufficient amount of energy to qualify as Type I, yet is actively destroying the very world from which it is deriving those $10^{16}$ Watts, its existence as a technical civilization would clearly be in great peril. Sustainment at Type I, let alone any prospect of advancement towards Type II, would be doubtful as the grim prospect of backsliding into a primal state grows. Such threats can arise from the reckless generation and usage of energy leading to environmental, biological and/or sociopolitical collapse and must be guarded against ever more robustly as humanity's technical prowess continues to outrun our capability to manage our increasingly complex society. Indeed, "The real problem of humanity is the following: We have Paleolithic emotions, medieval institutions, and godlike technology" [21]. This sobering insight, made by one of the great pioneers of evolutionary biology, stands as a stark reminder that the human species is still struggling with our troubled evolution in such places as Eastern Europe and the Middle East.

Another major and inevitable concern with the increasing development of nuclear energy are the dangers to all life on Earth posed by such a powerful resource, while trying to successfully avoid the Great Filter. Nuclear power is, by all means, a promising source of (essentially) clean energy that can meet our growing needs given it possesses the highest capacity factor, about 92.5%, in comparison to the other energy sources including non-renewables. *This basically means nuclear power plants are producing maximum power more than 93 percent of the time during the year* (U.S. Energy Information Administration, 2020). But the biggest concern in the course of nuclear power generation is in the safe handling and disposing of the byproduct hazardous nuclear wastes that can remain radioactive for hundreds, if not thousands, of years. The International Atomic Energy Agency suggests long term climate change must be taken into account while considering *Near Surface Disposal* of *Low Level* and *Intermediate Level Wastes* as this may affect their safe storage. But in considering the larger issue of nuclear waste disposal, we must also be aware of the major downside posed by soil degradation due to radioactive contamination. Radiation can affect the environment at any stage of the nuclear fuel cycle. Primarily, the processing of radioactive waste may lead to accidental release of radionuclides. *By a review of the inventory of fission products, important in case of accidental releases* [22], risks of contamination from key pollutants (e.g., radio-isotopes of Cesium) prevail at an average of two out of the four stages of the nuclear fuel cycle, with radio-isotopes of Strontium occurring across all four stages. *Although the stable forms of Strontium are not considered as toxic to plants, the adverse effect on plant development and growth is very often combined with its negative impact on the uptake of some nutrients, especially Calcium* [23]. In addition, corrosion products and generation of Technologically Enhanced Naturally Occurring Radioactive Material also act as significant soil contaminants. Analysis of soil properties, particularly those responsible for immobilization of radio-nutrients essential for controlling activity concentration in the biota, suggest decreases in pH value and clay content increase the risk of mobility of radionuclides in the soil. Deforestation and acid rain are the main reasons for removal of the clay content from soil and decline in soil pH. According to the Food and Agricultural Organization of the United Nations, between 2015 and 2020 the rate of deforestation was estimated at 10 million hectares per year, down from 16 million hectares per year in the 1990s. Worldwide, the area of primary forest has decreased by over 80 million hectares since 1990. If deforestation at this rate continues, we will soon be facing the major risk of oxygen reduction in soil along with greater risks of radioactive contamination. Additionally, *data from Landsat satellites* (Earth Observatory, NASA) suggests global warming has led to higher rates of forest fires, extending their impact to some hillsides and mountainsides which historically had incurred only rare incidents of such fires. *"Not only is the gas pedal on,"* says University of Colorado, Boulder, wildlife scientist Jennifer Balch, *"but the brakes are also off."* [24]. But what is even more concerning is that forest fires facilitate the spread of radionuclides over large distances by way of resuspension, especially in already contaminated areas like Chernobyl.

Thus, in concert with significantly increasing the growth rate of nuclear energy generation, we must also promote and conduct afforestation at a greater pace while adopting improved technology for even more secure disposal radioactive wastes, all while transitioning to cleaner forms of energy. Between 2002 and 2020, however, forest area in the US did not increase significantly. Furthermore, in developing countries like India access to renewable energy is much less as compared to the consumption of fossil fuels. This is mainly attributable to the lack of required technological growth and generally poor economic conditions of the populace, further slowing the transition from non-renewable to cleaner forms of energy.

In summary, for the entire world population to reach the status of a Kardashev Type I civilization we must develop and enable access to more advanced technology to all responsible nations while making renewable energy accessible to all parts of the world, facilitated by governments and private businesses. Only through the full realization of our mutual needs and with broad cooperation will humanity acquire the key to not only avoiding the Great Filter but continuing our ascent to Kardashev Type I, and beyond.

**Acknowledgement:** This work was supported by the Jet Propulsion Laboratory, California Institute of Technology, under contract with NASA. We acknowledge NASA ROSES Exoplanet Research Program for funding support.

**Data availability:** All data and software used for this study are submitted online as attachments. For additional questions regarding the data sharing, please contact the corresponding author at Jonathan.H.Jiang@jpl.nasa.gov.